\documentclass[aps,prl,amsmath,twocolumn]{revtex4}

\usepackage{appendix}
 
\newcommand{\R}{\mathcal{R}}
\newcommand{\E}{\mathcal{E}}

	\usepackage{amsmath}
	\usepackage{amsfonts,amsmath,amssymb}
  	\usepackage{amssymb}
  	\usepackage{float}
	\usepackage{makeidx}
	\usepackage{amsfonts}
	\usepackage[ansinew]{}
	\usepackage[usenames,dvipsnames]{pstricks}
	\usepackage{epsfig}

\usepackage{xcolor}

\newcommand{\be}{\begin{equation}}
\newcommand{\ee}{\end{equation}}
\newcommand{\bea}{\begin{eqnarray}}
\newcommand{\eea}{\end{eqnarray}}

\newcommand\SPD{\mathrel{\stackrel{\makebox[0pt]{\mbox{\normalfont\tiny (3)}}}{\Delta}}}

\newcommand{\dd}{{\mathrm{d}}}

\usepackage{xcolor}

\newcommand{\SA}[1]{{ #1}}

\makeindex

\begin{document}

\title{Effects of  the modification of gravity on the production of primordial black holes}
\author{Sergio Andr\'es Vallejo-Pe\~na${}^{2,4}$, Antonio Enea Romano${}^{1,2,3,4}$}

\affiliation{
${}^{1}$Theoretical Physics Department, CERN, CH-1211 Geneva 23, Switzerland\\
${}^{2}$ICRANet, Piazza della Repubblica 10, I--65122 Pescara, Italy \\
${}^{3}$Department of Physics \& Astronomy, Bishop's University\\
2600 College Street, Sherbrooke, Qu\'ebec, Canada J1M~1Z7
${}^{4}$Instituto de Fisica, Universidad de Antioquia, A.A.1226, Medellin, Colombia\\
}

\begin{abstract}
The enhancement of the spectrum of primordial comoving curvature perturbation $\R$ can induce the production of primordial black holes (PBH) which could account for part of present day dark matter.
As an example of the effects of the modification of gravity on the production of PBHs, we investigate the effects on the spectrum of $\R$ produced by the modification of gravity in the case of G-inflation, deriving the relation  between  the unitary  gauge curvature perturbation $\zeta$ and the comoving curvature perturbation $\R$, and identifying a background dependent enhancement  function  $\E$ which can induce large differences between  the two gauge  invariant variables. We use this relation  to derive  an equation for $\R$, showing for the presence of a momentum dependent effective sound speed (MESS), associated to the intrinsic entropy which can arise in modified gravity theories, in agreement with the model independent MESS approach to cosmological perturbations. 

When $\zeta$ is not constant in time it is different from $\R$, for example on sub-horizon scales, or in models exhibiting an anomalous super-horizon growth of $\zeta$, but since this growth cannot last indefinitely, eventually they will coincide.
We derive the general condition for super-horizon growth of $\zeta$, showing that  slow-roll violation is not necessary.
Since the abundance of PBHs  depends on the statistics of the peaks of the comoving density contrast, which is related to the spectrum of $\R$, it is important to take into account these effects on the PBHs abundance in  modified gravity theories.
\end{abstract}



\maketitle

\section{Introduction}
The study of primordial perturbations is fundamental in any cosmological model, since it allows to make predictions of the  conditions which provided the seeds for the anisotropies of the cosmic microwave background (CMB) radiation or for the process of structure formation.
Among the different theoretical scenarios proposed to explain the accelerated expansion of the Universe, Horndeski's theory \cite{Horndeski:1974wa} has received a lot of attention, both in the context of inflation and dark energy.

The calculation of the equation  for cosmological perturbations for these theories have been so far performed in the so called unitary gauge, also known as uniform field gauge. While the unitary gauge has some computational convenience in general relativity when only a scalar field is present, in general it is not directly related to observations, which depend on the comoving curvature perturbations $\R$. The production of PBHs \cite{Belotsky:2014kca,Carr1975,Khlopov:2008qy,Sasaki:2016jop,Josan:2009qn} is an example of phenomenon depending on $\R$ \cite{Sasaki:2018dmp} and not on the unitary gauge curvature perturbations $\zeta$.
Another example are the numerical codes developed for the solution of the Boltzman's equations in a perturbed Friedman-Lemaître-Robertson-Walker (FLRW) Universe, which are using equations in the synchronous gauge \cite{Bucher:1999re}, which for adiabatic perturbations 
coincides  approximately with the comoving gauge \cite{Romano:2015vxz},  justifying the use of the comoving slices gauge for early Universe calculations.

The comoving gauge can differ from the unitary gauge in modified gravity theories because the effective energy momentum tensor arising from the modification of gravity can produce some effective entropy  terms, which are absent in $K(X)$ theories, but are present in any more complicated Hordenski's theory.
The general form of the equation of curvature perturbation in comoving gauge $\R$ was derived in \cite{Romano:2018frb} assuming an arbitrary form of the total effective energy-stress tensor (EST), but no explicit calculation was given in the case of modified gravities.
In this letter we compute the general relation between $\R$ and $\zeta$ and use it to derive an equation for $\R$ for G-inflation, confirming the general form predicted in \cite{Romano:2018frb}, showing evidence of a momentum dependent effective sound speed (MESS).

One simple mechanism to produce PBHs in single field models is a violation of the slow-roll conditions \cite{Garcia-Bellido:2017mdw,Vallejo-Pena:2019lfo,Novikov:2016fzd}, which can induce a super-horizon growth of curvature perturbations, due to the growth of what would be a decaying mode during slow-roll \cite{Romano:2016gop,Romano:2016jlz}.
As an application  we use the gauge transformation between $\R$ and $\zeta$ to investigate the effects of the modification of gravity on the power spectrum of $\R$ in models violating slow-roll, such as for example ultra slow-roll G-inflation \cite{Hirano:2016gmv}, and its implications on the production of PBHs.

\section{G-inflation}
In G-inflation the scalar field $\Phi$ is minimally coupled to gravity according to the action \cite{Kobayashi:2010cm,Deffayet:2010qz} 
\be 
S= \int d^4x \sqrt{-g} \left(\frac{M_{Pl}^2}{2} R +  L (\Phi,X) \right) \, , \nonumber
\ee
where $X=-g^{\mu \nu}\partial_\mu \Phi \partial_\nu \Phi/2$, $R$ is the Ricci scalar and we use a system of units in which $c=\hbar=1$. The Lagrangian density of the scalar field corresponds to 
\be 
L(\Phi,X) = K(\Phi,X) + G(\Phi,X) \Box \Phi \, , \label{LKGB}
\ee
where $K$ and $G$ are arbitrary functions. 
The corresponding effective stress-energy-momentum tensor (EST) is given by
\begin{align}
T_{\mu \nu} = L_{,X} \nabla_{\mu} \Phi \nabla_{\nu} \Phi + P_{\Phi} g_{\mu \nu}  + \nabla_{\mu} \Phi \nabla_{\nu} G + \nabla_{\nu} \Phi \nabla_{\mu} G \, , \label{EMKGB} \end{align}

where 
\begin{align}
L_{,X}= \partial_{X}L =K_X(\Phi,X)+G_X(\Phi,X) \square \Phi \, ,  \\
P_{\Phi}=L - \nabla_{\mu}\left( G \nabla^{\mu}\Phi \right) =K - g^{\mu \nu} \nabla_{\mu}\Phi \nabla_{\nu}G \, .
\end{align}  
\section{The perturbed effective energy-stress-momentum tensor}
The most general scalar perturbations with respect to a flat FLRW background can be written as 
\begin{align}
    ds^2=a^2 \Big\{& -(1+2A)d\tau^2+2\partial_iB dx^id\tau+   \nonumber \\ \quad {} & + \left[ \delta_{ij}(1-2C)+2\partial_i\partial_jE\right]dx^idx^j \Big\} \,.  \label{pmetric}
\end{align}

For the  decomposition of the scalar field and the EST into their background and perturbation parts we use the notation
\begin{align} 
\Phi(x^\mu) &=\phi(\tau)+\delta\phi(x^\mu) \, \\  \label{pfield} 
T^{\mu}{}_{\nu} &= \overline{T}^{\mu}{}_{\nu} + \delta T ^{\mu}{}_{\nu} \,.
\end{align}
The background components of the EST are 
\begin{align}
\overline{T}^{0}{}_{0}=& -\overline{\rho} = K(\phi,\chi) + \frac{3 \mathcal{H} \phi'{}^3}{a^4} G_{\chi}(\phi,\chi)+  \nonumber \\ \quad {} & - \frac{\phi'{}^2}{a^2} \left[ K_{\chi}(\phi,\chi) + G_{\phi}(\phi,\chi)\right]  \, , \\ \label{bT00}
\overline{T}^{0}{}_{i}=& \overline{T}^{i}{}_{0}= 0 \, ,   \\
\overline{T}^{i}{}_{j}=& \delta ^{i}{}_{j} \overline{P} \, , \nonumber \\ \overline{P} =&    K(\phi,\chi)  - \frac{\mathcal{H} \phi'^3}{a^4} G_{\chi}(\phi,\chi)+  \nonumber \\  \quad{} & +  \frac{\phi'{}^2}{a^2} \left[ G_{\phi}(\phi,\chi)+ \frac{\phi''}{a^2}G_{\chi}(\phi,\chi) \right]   \, , \label{bTij}
\end{align}
where the primes stand for derivatives with respect to $\tau$, $\chi$ is given by $\chi=\frac{\phi'^2}{2a^2}$, and the subscripts $\phi$ and $\chi$ denote partial derivatives with respect to these quantities, i.e. $G_{\phi}(\phi,\chi)=\partial_{\phi}G(\phi,\chi)$ and $G_{\chi}(\phi,\chi)=\partial_{\chi}G(\phi,\chi)$.
In order to define the comoving slices gauge we need this component of the perturbed  EST 
\begin{align}
\delta T ^{0}{}_{i} =& - \left( K_{\chi} + 2 G_{\phi} - \frac{3\mathcal{H}\phi'}{a^2}G_{\chi}\right) \frac{\phi'{}^2}{a^2}\partial_i\delta\phi+ \nonumber \\ \quad {} & -\frac{\phi'{}^2}{a^4}G_{\chi} \partial_i \left( \delta\phi' - \phi' A \right) \, ,\label{dT0i}
\end{align}
where $\mathcal{H}=a'/a$.
The remaining components of the perturbed EST are not relevant to the computations done in this letter, and we will give them in a future work. 
Under a gauge transformation of the form $(\tau,x^i)\to (\tau+\delta \tau,x^i+\delta x,^i)$ the perturbations $\delta \phi$, $A$, $B$, $C$, and $E$ transform according to \cite{Kodama:1985bj}
\begin{align}
    \delta \phi &\to \delta\phi-\phi'\delta \tau \, , \label{gtphi} \\ 
    A &\to A- \mathcal{H} \delta \tau - \delta \tau' \, , \label{gtA} \\
    B &\to B + \delta \tau - \delta x' \, , \label{gtB} \\ 
    C &\to C + \mathcal{H} \delta \tau \, , \label{gtC} \\
    E &\to E - \delta x \, . \label{gtE}
\end{align}

\section{Evolution of curvature perturbations in the unitary gauge}
 In single scalar field models the unitary gauge is defined by the condition $\delta\phi_{u}=0$. From  the gauge transformation in eq.(\ref{gtphi}) we can see that the time translation $\delta \tau_{u}$ necessary to go to the unitary gauge is given by
\be
 \delta \tau_{u} = \frac{\delta \phi}{\phi'} \, . 
\ee
Using eq.(\ref{gtC}) we can compute the curvature perturbation in the unitary gauge $\zeta$ 
\begin{align}
    \zeta \equiv  -C_{u} =
    -C - \mathcal{H} \delta \tau_{u} 
    = - C - \mathcal{H} \frac{\delta \phi}{\phi'}  \, . \label{defzeta}
\end{align}
which is by construction  gauge invariant.
We can also define other gauge invariant quantities such as the unitary gauge lapse function 
\begin{align}
    A_{u} \equiv & A- \mathcal{H}\delta\tau_{u}-\delta\tau_{u}'
          =A- \mathcal{H}\frac{\delta \phi}{\phi'}- \left(\frac{\delta \phi}{\phi'}\right)' \, .
\end{align}

The second order action for $\zeta$ in Horndeski's theories  was  computed in \cite{Kobayashi:2011nu}
\begin{align}
    S^{(2)}_{\zeta} =& \int dt d^3x a^3 \left[ \mathcal{G}_S \dot{\zeta}^2 - \frac{\mathcal{F}_S}{a^2} \left(\partial_i \zeta\right)^2 \right] \, , 
\end{align}
where $\mathcal{G}_S$ and $\mathcal{F}_S$ are  functions of $K(\phi,\chi)$ and $G(\phi,\chi)$ and their derivatives. 
The Lagrange equations for this action give the  equation of motion of $\zeta$ 
\be
\zeta''+ \left( 2\mathcal{H} + \frac{\mathcal{G}_S'}{\mathcal{G}_S} \right) \zeta' - c_s^2 \SPD \zeta = 0 \, , 
\ee
where $c_s^2(\tau)=\mathcal{F}_S/\mathcal{G}_S$.  
For the Fourier transform of the above equation we use the notation
\be
\zeta_k''+ \left( 2\mathcal{H} + \frac{\mathcal{G}_S'}{\mathcal{G}_S} \right) \zeta_k' + c_s^2 k^2 \zeta_k = 0 \, . \label{eqzeta} 
\ee
\section{Enhancement of curvature perturbations}
As already observed for  comoving curvature perturbation $\R$ in general relativity for standard kinetic term single field models  \cite{Saito:2008em},  a temporary violation of slow-roll conditions can lead to the anomalous growth of what would normally be a decaying mode. A similar mechanism can induce the growth of $\zeta$, as we will show in this section.
We can re-write eq.(\ref{eqzeta}) in the form
\begin{align}
    \frac{\dd}{\dd a}\left( a^3 \mathcal{H} \mathcal{G}_S   \frac{\dd \zeta_k}{\dd a} \right) + a \mathcal{F}_S \frac{k^2}{\mathcal{H}} \zeta_k &= 0 \, , \label{eqzetaM}
\end{align}
from which it is possible to find a super-horizon scale solution of the form
\bea
  \zeta_k &=& A + B \int \frac{\dd a}{a} f  \, ,  \label{zetaf} \\
  f &=& \frac{1}{a^2\mathcal{H}\mathcal{G}_S} \,,
  \label{f}
\eea

where $A$ and $B$ are constants. For standard slow-roll models the function $f$ decreases as the scale factor increases, implying that $\zeta$ tend to a constant value, i.e. the second term in eq.(\ref{zetaf}) is a decaying mode. If the function $f$ is a growing function of $a$ then the second term in eq.(\ref{zetaf}) becomes a growing mode, and there can be a super-horizon growth.
It follows that the general condition for super-horizon growth of $\zeta_k$ is then 
\be
\frac{\dd f}{\dd a} \geq 0 \, , \label{gc}
\ee
or equivalently
\begin{equation}
    \frac{\dd f}{\dd a} =\frac{1}{a'} \frac{\dd f}{\dd \tau} = \frac{1}{a \mathcal{H}} f' \geq 0 \, . 
\end{equation}
During inflation $a\mathcal{H}>0$ and this condition reduces to
\begin{equation}
    f'\geq0 \, . \label{fdg}
\end{equation}

In the case of a minimally coupled single scalar field the unitary gauge and the comoving gauge coincide, and the general condition given above takes the form
\cite{Saito:2008em}
\be
3-\epsilon+\eta\leq0
\ee
where the slow-roll parameters are defined according to
\be
    \epsilon \equiv -\frac{a}{\mathcal{H}^2} \left( \frac{ \mathcal{H}}{a}\right)' = \frac{ a^2 ( \overline{\rho}+\overline{P} ) }{2 M_{Pl}^2 \mathcal{H}^2} \quad , \quad  
    \eta \equiv \frac{\epsilon'}{\epsilon \mathcal{H}} \, . \label{eta}
\ee

In G-Inflation the condition given in eq.(\ref{fdg}) implies that
\begin{align}
    f' &= \frac{\dd}{\dd \tau}\left(\frac{1}{a^2\mathcal{H}\mathcal{G}_S}\right)= \frac{3-\epsilon+\mathcal{G}'_S/\mathcal{H}\mathcal{G}_S}{a^2\mathcal{G}_S}=\frac{\gamma}{\delta}\leq0 \label{fdsr}\, ,
\end{align}
which gives the general condition for super-horizon growth in an expanding Universe. For a contracting Universe the inequality would be inverted.

As can be seen from the above equation the super-horizon growth can be achieved in different cases, corresponding to $\gamma$ and $\delta$ having opposite signs,
contrary to what happened for the standard kinetic term single field scenario, in which $\delta$ sign is fixed.
Note also that contrary to standard kinetic term single field models, the super-horizon growth does not depend only on the slow-roll parameters, implying that it can occur also during slow-roll.

The anomalous super-horizon growth of $\zeta$, and consequently of $\R$, can increase the abundance of PBHs, since it affects the statistics of the density perturbations peaks which can seed the PBHs. We will discuss this  in more details in the following section.
\section{Comoving slices gauge in G-inflation }

The comoving slices gauge is defined by the condition $\delta T ^{0}{}_{i}=0$. In  G-inflation, combing eqs.(\ref{gtphi}-\ref{gtA}) with eq.(\ref{dT0i}) we have that under an infinitesimal time translation 
\be
\delta T ^{0}{}_{i} \to \delta T ^{0}{}_{i} + \partial_i \left(\frac{\phi'{}^2}{a^4} D \delta \tau \right) \, , 
\ee
where
\begin{align}
    D = & a^2(2G_{\phi}+K_{\chi}) + G_{\chi} (-4\mathcal{H} \phi'+\phi'') \, ,
\end{align}
from which we get the time translation $\delta \tau_c$ required to go to the comoving slices gauge 
\begin{align}
\delta \tau_c = \frac{1}{\phi'D} \Big[ & - \phi' G_{\chi} (3\mathcal{H} \delta\phi+\phi'A-\delta\phi') + \nonumber  \\  \quad {} & + a^2(2G_{\phi}+K_{\chi})\delta\phi  \Big] \, . \label{deltatauc}
\end{align}

Note that in the particular case in which $G$ does not depend explicitly on $\chi$, i.e. $G(\phi,\chi)=G(\phi)$ the above transformation reduces to
\be 
\delta \tau_c = \frac{\delta \phi}{\phi'} \, , \label{deltataucp}
\ee
and the comoving gauge coincides with the unitary gauge, since in this case the system is equivalent  to a $K(X)$ theory \cite{Romano:2016jlz,Garriga:1999vw}.

The comoving curvature perturbation $\R$ is then defined as
\begin{align}
    \R \equiv & -C_c = -C - \mathcal{H} \delta\tau_c \, . \label{defR}
\end{align}
Our goal is to derive \SA{the relation between $\zeta$ and $\R$}, and we can achieve this by performing the gauge transformation between the unitary and comoving slices gauge. \SA{We can also derive the equation of motion of $\R$ from eq.(\ref{eqzeta}) using this relation, as shown in  the appendix.}

Using  the general gauge transformation defined  in eq.(\ref{deltatauc}), when $\delta\phi=0$ and $A=A_u$, we get
\be
\delta\tau_{uc} =- \frac{\phi'G_{\chi}}{D} A_{u} \,,
\ee
from which we obtain
\begin{align}
    \R =& \zeta + \mathcal{H}\frac{\phi'G_{\chi}}{D} A_{u} \, . \label{RzetaAun}
\end{align}
The gauge invariant variable $A_u$ can be expressed in terms of $\zeta$ using  the perturbed Einstein's equation $\delta G^{0}{}_{i} =  \delta T^{0}{}_{i}/M_{Pl}^2 $ in the unitary gauge, which using  eq.(\ref{dT0i}) gives
\begin{align}
    -\zeta'+\mathcal{H}A_{u} =& - \frac{\phi'{}^3 G_{\chi}}{2M_{Pl}^2a^2} A_{u} \, . \label{zetapAun}
\end{align}{}
We can then combine eq.(\ref{RzetaAun}) and eq.(\ref{zetapAun}) to obtain \SA{the} relation between $\R$ and $\zeta$ only
\bea
\R&=&  \zeta + \mathcal{H}\frac{\phi'G_{\chi}}{D} \left( \frac{\phi'{}^3 G_{\chi}}{2M_{Pl}^2a^2}+ \mathcal{H}\right)^{-1} \zeta'  \nonumber \\
&=&\zeta+\E(\tau) \zeta' \, . \label{Rzeta}
\eea
where we have defined the enhancement factor $\E(\tau)$, a quantity depending only on the background, which can induce a significant difference between the curvature perturbations on comoving  and  uniform field slices.
The relation between the power spectrum of $\zeta$ and $\R$ is then given by

\be
P_{\R}=\frac{k^3}{2\pi^2}|\R_k|^2=P_{\zeta}+\frac{k^3}{2\pi^2} \Delta \label{PR}
\ee
where
\be
\Delta=\Big[ \E \zeta^*\zeta'+\E^*\zeta'^*(\zeta+\E \zeta') \Big] 
\ee

Note that the above relations are valid on any scale, since they are just based on gauge transformations, without assuming any sub or super horizon limit.
This implies that the spectra of $\R$ and $\zeta$ could be different due to a change in the evolution of both sub-horizon and super-horizon modes during the time interval when $\E(\eta)$ is large. On sub-horizon scales the effect is always present, since $\zeta$ is oscillating and $\zeta' \neq 0$, while for super-horizon scales the effect could be suppressed if $\zeta \approx 0$, but even for models conserving $\zeta$ there could be an effect, since the freezing does not happen immediately after horizon crossing. We will discuss later the implication on the production of PBHs.
\section{Conservation of $\R$ and $\zeta$ }
From eq.(\ref{Rzeta}) we can reach the important conclusion that 
\be
\zeta=const \Rightarrow \zeta=\R=const \, ; \label{zetacon}
\ee
however the opposite is not true, i.e.
\be
\R=const \nRightarrow \zeta=const \, ,
\ee
which can have important implications for conservation laws of $\R$ and non-Gaussianity consistency conditions \cite{Romano:2016gop}. 
As explained previously, $\R$  is the quantity related to observations, so it would be inconsistent to infer constraints on $\zeta$ from CMB observations for example, since the latter depend on $\R$.
From a theoretical point of view the models approximately conserving  $\zeta$ on super-horizon scales may  be incompatible with observations for large enhancement functions $\E(\tau)$, because $\R$ could be not conserved, implying for example a violation of the non-Gaussianity consistency condition or a miss-estimation of PBHs abundance.

Nevertheless it should be noted that the super-horizon growth of perturbations cannot last indefinitely, or the entire perturbative treatment of the problem would breakdown, leading to inhomogeneities much larger than those imprinted in the CMB for example. For this reason it is expected that for any model compatible with observations the super-horizon growth of $\zeta$ should be only temporary, and according to eq.(\ref{zetacon}), at some time after horizon crossing  $\zeta \approx \R$.
This simplifies the calculation of  $\R$, whose evolution can be then traced during and after reheating, too, contrary to $\zeta$. In fact the  equation of motion for $\R$, which we give in the appendix, is rather  complicated compared to that for $\zeta$.

The only exception to this argument could be very small scales $\zeta$ modes which leave the horizon very late, and whose super-horizon growth could continue until horizon re-enter, without affecting the validity of the perturbative treatment of the problem. For these small scale modes the difference between $\R$ and $\zeta$ could be important, but it would still be  computationally convenient to solve the equation for $\zeta$ and then obtain $\R$ using the gauge transformation given in eq.(\ref{Rzeta}).

\section{Production of primordial black holes}
The super-horizon growth of $\R_k$ could  produce primordial black holes which could possibly account for part of dark matter \cite{Carr:2009jm,Carr:2020gox,Yokoyama:1995ex,Sasaki:2018dmp,Garcia-Bellido:2017mdw,GarciaBellido:1996qt,Ivanov:1994pa,Chapline1975,Belotsky:2018wph,Belotsky:2014kca,Khlopov:2008qy} and produce gravitational waves (GW) detectable with future GW detectors such as LISA \cite{Sasaki:2018dmp,Sasaki:2016jop}. In this session we will show how to obtain some approximate estimation of the effetcs of the modification of gravity on the PBH production, without considering any specific model, leaving this to a future work.

The mass $M$ of PBHs produced by the mode $\R_k$ re-entering  the horizon during the radiation domination can be approoximated as \cite{Sasaki:2018dmp} \begin{equation}
    M = \gamma M_{H}\Bigr|_{F} \, ,
\end{equation}
where $\gamma \approx 0.2$ is a correction factor, and $M_{H}\Bigr|_{F}$ is the horizon mass $M_H\equiv(4\pi/3)\overline{\rho} (a\mathcal{H})^{-3} $ at the time of PBH formation, corresponding to the horizon crossing time  
\begin{equation}
    k = (a^2 \mathcal{H})\Bigr|_{F} \, .
\end{equation}
Note the above is just a rough estimation, and a more accurate treatment would involve the use of a scaling relation\cite{Yokoyama:1998xd,Niemeyer:1997mt}.  

The present time fraction $f_{PBH}$ of PBHs of mass M
against the total dark matter component can then be approximated as \cite{Sasaki:2018dmp}
\begin{align}
    f= 2.7 \times 10^8 \left(\frac{\gamma}{0.2}  \right)^{1/2} \left(\frac{g_{*F}}{106.75}  \right)^{-1/4} 
     \left(\frac{M}{M_{\odot}}  \right)^{-1/2} \beta \, \nonumber ,
\end{align}
where $g_{*F}$ is the number of relativistic degrees of freedom at formation, The quantity $\beta$ is  the energy density fraction of PBHs at formation time
\begin{equation}
    \beta \equiv \frac{\overline{\rho}_{PBH}}{\overline{\rho}} \Bigr|_{F} \, ,
\end{equation}
which can be written in terms of the probability of  the density contrast $P(\delta)$ as \cite{Carr1975,Green:2004wb}
\begin{equation}
    \beta(M) = \gamma \int_{\delta_{t}}^{1} P(\delta) \dd \delta \, ,
\end{equation}
where  $\delta_t$ is the threshold for PBH formation. Assuming the density perturbations follow a Gaussian distribution $\beta$ is given by
\begin{equation}
    \beta(M) \approx \frac{\gamma}{\sqrt{2\pi}\nu(M)} \exp\left[ -\frac{\nu(M)^2}{2}\right] \, , \label{betaM}
\end{equation}
where $\nu(M)\equiv \delta_t/\sigma(M)$, and $\sigma(M)$ is an estimation of the standard deviation  of the density contrast on scale $R$ from the variance
\begin{align}
    \sigma^2(M) &= \int \dd \ln k W^2(k R)\mathcal{P}_{\delta}(k) \nonumber \\ &= \int \dd \ln k W^2(k R) \left( \frac{16}{81}\right)(k R)^4\mathcal{P}_{\R}(k) \, , 
\end{align}
where $W(k R)$ is a window function smoothing over the comoving scale $R(M)=(a^2 \mathcal{H})^{-1}\Bigr|_{F}=2 G M/a_{F}\gamma^{-1}$, and the relation between $\delta$ and $\R$ has been used in the second equality.
It should be mentioned that eq.(\ref{betaM}) can be used as a guideline, but more accurate calculations would involve the use of the results of numerical simulations \cite{Nakama:2013ica,Shibata:1999zs}. The choice of the window function could also affect \cite{Ando:2018qdb,Tokeshi:2020tjq} the results  of the calculation.

Our aim here is not make an accurate estimation of the PBHs abundance for a specific model, but to show why in general it can be impacted by the modification of gravity, and the approximations adopted so far are enough to serve this general purpose.
According to the equations above, the PBH fraction $\beta$ is affected by the power spectrum  of $\R$  since this can increase the standard deviation of the density field $\sigma(M)$.
Note that the above  approximations to estimate the PBHs abundance can  receive  important corrections depending on the shape of power spectrum, on  non- gaussianity, and non-linear statistics \cite{Germani:2018jgr,Atal:2018neu,Germani:2019zez}. 
Due to the importance of all these different effects it is difficult to find a general model independent analytical formula to estimate the PBHs abundance for a generic G-inflation theory, but any enhancement of the power spectrum is expected, according to eq.(\ref{PR}), to affect the probability  of production of PBHs.
Beside this, numerical relativity simulations of the PBHs formation are based on general relativity, so the effects of the modification of gravity on the process of gravitational collapse are at the moment not fully understood and would require investigations beyond the scope of this paper \cite{Chen:2019qmt}.


At the end of  its  anomalous super-horizon growth, $\zeta$  will coincide with $\R$, and the consequent  enhancement of the spectrum will lead to an increased PBH  abundance.
Contrary to what happens for standard kinetic term single field models in general relativity \cite{Saito:2008em}, in the case of G-inflation this power spectrum enhancement can be achieved also during slow-roll, as long as the condition in eq.(\ref{fdsr}) is satisfied, which can be attained by an appropriate choice of the function $\mathcal{G}_S$.
We expect a similar behavior for more complex modified gravity theories as well.


\section{Conclusions}
We have computed the effective energy-stress-tensor for G-inflation theories in the comoving slices gauge and have used it to derive a general relation between the unitary gauge curvature $\zeta$ and the comoving curvature perturbation $\R$, involving an enhancement function which depends on the evolution of the background, and which can cause a large difference between the two gauge invariant quantities. We have then derived an equation for $\R$ and used it to determine its super-horizon behavior.
The equation shows the presence of a momentum effective sound speed, due to intrinsic entropy, in agreement withe MESS approach to cosmological perturbations.

When $\zeta$ is not constant in time it  differs from $\R$, for example on sub-horizon scales, or in models exhibiting an anomalous super-horizon growth of $\zeta$, but since this growth cannot last indefinitely, eventually they will coincide.
We have derived the general condition for super-horizon growth of $\zeta$, showing that slow-roll violation is not necessary, and discussed how the the enhancement of the spectrum of $\R$ can affect the PBH abundance. 

We expect similar results to hold for other modified gravity theories such as other Horndeski's theories \cite{Horndeski:1974wa}, since also for these theories there can be effective entropy or anisotropy terms which can modify the evolution of curvature perturbations. 
In the future it will be interesting to extend this study to other modified gravity theories or to multi-fields systems, and to use observations to constraints the different types of theories.
It would also be important to perform numerical simulations of the PBHs formation taking into account the non perturbative effects of the modification of gravity on the process of black hole formation.





\begin{appendices}
\section{Evolution of $\R$ in G-inflation}
We can  use eq.(\ref{eqzeta}) and eq.(\ref{Rzeta}) to derive the equation for $\R$ in Fourier space
\begin{equation}
    \R_k'' + \alpha_k(\tau) \R_k' + \beta_k(\tau) k^2 \R_k =0 \, , \label{eqR}
\end{equation}
with the coefficients $\alpha_k$ and $\beta_k$ given by
\allowdisplaybreaks
\begin{align}
    \alpha_k =& \frac{1}{\mathcal{D}_k} \Bigg\{ \E k^2 c_s \mathcal{G}_S \bigg[-2 \E \mathcal{G}_S c_s'+c_s \Big(\E \left(2 \mathcal{H} \mathcal{G}_S+\mathcal{G}_S'\right)+ \nonumber \\ \quad {} & -2 \E' \mathcal{G}_S\Big)\bigg]+ \mathcal{G}_S^2 \bigg[-\E''+\mathcal{H} \left(4 \E'+2\right) -4 \E \mathcal{H}^2+ \nonumber \\  \quad & +2 \E \mathcal{H}'\bigg]+ \mathcal{G}_S \bigg[\left(2 \E'-4 \E \mathcal{H}+1\right) \mathcal{G}_S' +\E \mathcal{G}_S''\bigg] + \nonumber \\ \quad & -2 \E \mathcal{G}_S'{}^2 \Bigg\} \, , \label{alpha} \\ 
    \beta_k =& \frac{1}{\mathcal{D}_k} \Bigg\{ \E^2 k^2 c_s^4 \mathcal{G}_S^2 -\E^2 c_s^2 \mathcal{G}_S'{}^2 + c_s \mathcal{G}_S^2 \bigg[2 \E c_s' \Big(\E' + \nonumber \\ \quad &-2 \E \mathcal{H}+1\Big)+c_s \Big(2 \E^2 \mathcal{H}'+2 \E'{}^2+3 \E' -\E \big(\E''+ \nonumber \\ \quad &+2
   \mathcal{H} \left(\E'+1\right)\big)+1\Big)\bigg]+  \E c_s \mathcal{G}_S \bigg[-2 \E c_s' \mathcal{G}_S'+ \nonumber \\ \quad &  -c_s \left(\left(\E'+1\right) \mathcal{G}_S'-\E \mathcal{G}_S''\right)\bigg] \Bigg\} \, , \label{beta}
\end{align}
where 
\begin{align}
   \mathcal{D}_k = & \mathcal{G}_S \left(\mathcal{G}_S \left(\E^2 k^2 c_s^2+\E'-2 \E \mathcal{H}+1\right)-\E \mathcal{G}_S'\right) \, .
\end{align}
Note that, contrary to the K-inflation case, the coefficient of the Laplacian and that of the first time derivative are momentum dependent, while in the unitary gauge they are  only time dependent. This difference is related to the presence of intrinsic entropy as we will discuss in more details in the following section.
\section{Momentum dependent effective speed}
The equation derived in the previous section is in agreement with the general model independent result obtained in \cite{Romano:2018frb} 
\be
\R''_k + \frac{(\tilde{z}_k^2)'}{\tilde{z}_k^2} \R'_k + \tilde{v}_k^2 k^2 \R_k = 0 \, \quad,\quad \tilde{z}_k^2=\epsilon a^2/\tilde{v}_k^2
\label{MESSeq}
\ee
and  shows the presence of a momentum dependent effective sound speed (MESS), which is in fact expected to arise in modified gravity theories. 
The general model independent definition of the MESS is 
 \be
\tilde{v}_k^2 \equiv \frac{\tilde{\delta P_c}}{\tilde{\delta \rho_c}} \, ,
 \label{MESS}
 \ee
where $\tilde{\delta P_c}$ and $\tilde{\delta P_c}$ are the Fourier transform of the pressure and energy density perturbations in the comoving slices gauge. The momentum dependency comes from the presence of an intrinsic non adiabatic component of the comoving pressure perturbations of the effective EST of modified gravity theories, while in multi-fields systems \cite{Romano:2020oov} it is related to the entropy associated  to the presence of different degrees of freedom.
 Instead of using the gauge transformation, an alternative  approach for the calculation of the equation for $\R$  could have consisted in computing the MESS according to eq.(\ref{MESS}), and then replacing into the general eq.(\ref{MESSeq}).
 The MESS is an effective quantity which can be useful in  model independent analysis, and can for example explain anomalies of the CMB \cite{Rodrguez:2020hot}, but we will study these effects for G-inflation in a separate work.
\section{Equation of $\R$ in K-inflation}
In K-inflation $G=0$, implying that the unitary and comoving slices gauge coincide, i.e. $\R=\zeta$, and $\E=0$, which  replaced into eqs.(\ref{alpha}-\ref{beta}) give
\begin{align}
\alpha_k &= \left( 2\mathcal{H} + \frac{\mathcal{G}_S'}{\mathcal{G}_S} \right) =  \frac{(a^2\mathcal{G}_S)'}{a^2\mathcal{G}_S} \, , \\
\beta_k &= c_s^2 \, ,
\end{align}
as we were expecting from eq.(\ref{eqzeta}). Let us now compute this coefficients in order to show that eq.(\ref{eqR}) reduces to the well known equation in K-inflation models. In these models we have
\begin{align}
\mathcal{G}_S &= \frac{a^2\chi \overline{\rho}_{\chi}}{\mathcal{H}^2} \, , \label{GcsK} \\
\mathcal{F}_S &= -\frac{M_{Pl}^2(\mathcal{H}'-\mathcal{H}^2 )}{a^2\mathcal{H}^2}  \, ,\label{FcsK}
\end{align}
where 
\begin{equation}
   \overline{\rho}_{\chi} =  K_{\chi}(\phi,\chi) + 2 \chi K_{\chi\chi}(\phi,\chi)  \, . 
\end{equation}
After combining eqs.(\ref{GcsK}-\ref{FcsK}) with the background equation
\begin{equation}
    \frac{1}{a}\left( \frac{ \mathcal{H}}{a}\right)' = - \frac{\overline{\rho}+\overline{P}}{2 M_{Pl}^2} \, , \label{eqba}
\end{equation}
we obtain
\begin{equation}
\beta_k = c_s^2 = \frac{\overline{\rho}+\overline{P}}{2\chi\overline{\rho}_{\chi}} \, , \label{betaK}
\end{equation}
which coincides with the sound speed defined in K-inflation \cite{Garriga:1999vw}. Combining eq.(\ref{betaK}) and eq.(\ref{GcsK}) with the definition of the slow-roll parameter $\epsilon$ we find 
\begin{equation}
    z^2 = \frac{2 a^2 \epsilon}{c_s^2} = \frac{2 a^4 \chi \overline{\rho}_{\chi}}{M_{Pl}^2\mathcal{H}^2} = \frac{2 a^2 \mathcal{G}_S}{M_{Pl}^2} \, ,
\end{equation}
which implies 
\begin{equation}
   \alpha_k = \frac{(z^2)'}{z^2} \, .  \label{alphaK}  
\end{equation}
Thus, in the case of K-inflation eq.(\ref{eqR}) reduces to the well known Sasaki-Mukhanov equation 
\begin{equation}
    \R_k'' + \frac{(z^2)'}{z^2} \R_k' + c_s^2 k^2 \R_k =0 \, . \label{eqRK}
\end{equation}
\section{Super-horizon conservation of $\R$}
On super-horizon scales, assuming the gradient terms can be neglected,
 according to eq.(\ref{alpha}) $\alpha_k$ becomes a function of time only, which we denote as $\alpha$ 
\begin{align}
    \alpha =& \frac{1}{\mathcal{D}} \Bigg\{\mathcal{G}_S^2 \bigg[-\E''+\mathcal{H} \left(4 \E'+2\right) -4 \E \mathcal{H}^2 +2 \E \mathcal{H}'\bigg]+ \nonumber \\  \quad &+ \mathcal{G}_S \bigg[\left(2 \E'-4 \E \mathcal{H}+1\right) \mathcal{G}_S' +\E \mathcal{G}_S''\bigg] -2 \E \mathcal{G}_S'{}^2 \Bigg\} \, , \label{alphash}
\end{align}
where 
\begin{align}
   \mathcal{D} = & \mathcal{G}_S \left(\mathcal{G}_S \left(\E'-2 \E \mathcal{H}+1\right)-\E \mathcal{G}_S'\right) \, .
\end{align}
We can also re-write eq.(\ref{eqR}) on super-horizon scales as
\be
\left(\tilde{z}^2 \R_k' \right)'\approx 0 \, \label{Rcon} ,
\ee
where we have defined $(\tilde{z}^2)'/\tilde{z}^2 \equiv \alpha$, 
which implies that the conserved quantity is not $\R_k$ but $\R_k' \tilde{z}^2$.
Depending on the behavior of $\tilde{z}^2$, $\R_k$ may be conserved or not, implying a possible violation of the non-Gaussianity consistency condition \cite{Romano:2016gop}. The definition of $\tilde{z}$ implies
\begin{equation}
    \tilde{z}^2 \propto \exp \left( \int \dd \tilde{\tau} \alpha  \right) \, , \label{zt2a}
\end{equation}
and integrating  eq.(\ref{Rcon}) we can obtain the super-horizon behavior of $\R_k$ 
\be
\R_k \propto \int \frac{\dd \tau}{\tilde{z}^2} \propto \int \dd \tau \exp \left(-2 \int \dd \tilde{\tau} \alpha  \right) \, , 
\ee
implying that  $\R_k$ can increase when $\tilde{z}^2$ is decreasing.
This is consistent with eq.(\ref{Rzeta}), since  the enhancement function $\E(\tau)$ can induce a growth of $\R$. 
This work was partially supported by the Sostenibilidad program of UDEA.
\end{appendices}

\section*{Acknowledgments}
We thank Atsushi Naruko and Misao Sasaki for useful discussions and correspondence regarding the equation for $\R$, and  Ilia Musco and Cristiano Germani for discussions regarding the PBH formation.

\bibliographystyle{h-physrev4}
\bibliography{mybib}
\end{document}